\documentclass[prl,twocolumn,tightenlines,lengthcheck,showpacs,showkeys,preprintnumbers,superscriptaddress]{revtex4}

\usepackage{amsmath,amssymb,amsfonts,graphicx,dcolumn}

\newcommand{\bfr}{{\bf r}}

\newcommand{\bfq}{{\bf q}}

\def\la{\langle}
\def\ra{\rangle}

\begin{document}

\preprint{NT@UW-10-19}

\title{Third Zemach Moment of the Proton}

\author{Ian C. Clo\"{e}t}
\affiliation{Department of Physics, University of Washington, Seattle, WA 98195-1560}

\author{Gerald A. Miller}
\affiliation{Department of Physics, University of Washington, Seattle, WA 98195-1560}


\begin{abstract}
Modern electron scattering experiments have determined the proton electric form
factor, $G_{Ep}(Q^2)$, to high precision. We utilize this data, represented by
the different empirical form factor parametrizations, to compute the third Zemach moment of
the proton charge distribution. We find that existing data rule out a value of the
third Zemach moment large enough to explain the current puzzle with
the proton charge radius, determined from the Lamb shift in muonic hydrogen.
This is in contrast with the recent paper of De R\'{u}jula. We also demonstrate that
the size of the third Zemach moment is largely governed by the fourth moment of the 
conventional charge distributions, $\la r^4 \ra$, which enables us to obtain a rigorous
upper bound on the magnitude of the proton's third Zemach moment.
\end{abstract}


\maketitle


Pohl \textit{et al.} recently reported a very precise measurement of
the Lamb shift in muonic hydrogen, that leads to a proton radius of~\cite{pohl}
\begin{align}
\langle r_p^2\rangle^{1/2} = 0.84184 \pm 0.00067\, \text{fm}, 
\end{align}
which is five standard deviations away from the CODATA compilation 
of~\cite{CODATA}
\begin{align}
\langle r_p^2\rangle^{1/2} = 0.8768 \pm 0.0069\, \text{fm}.
\end{align}
There has been much speculation about the origin and possible consequences of the
puzzling difference between the results of muonic and electronic hydrogen.

\begin{figure}[tbp]
\centering\includegraphics[width=\columnwidth,clip=true,angle=0]{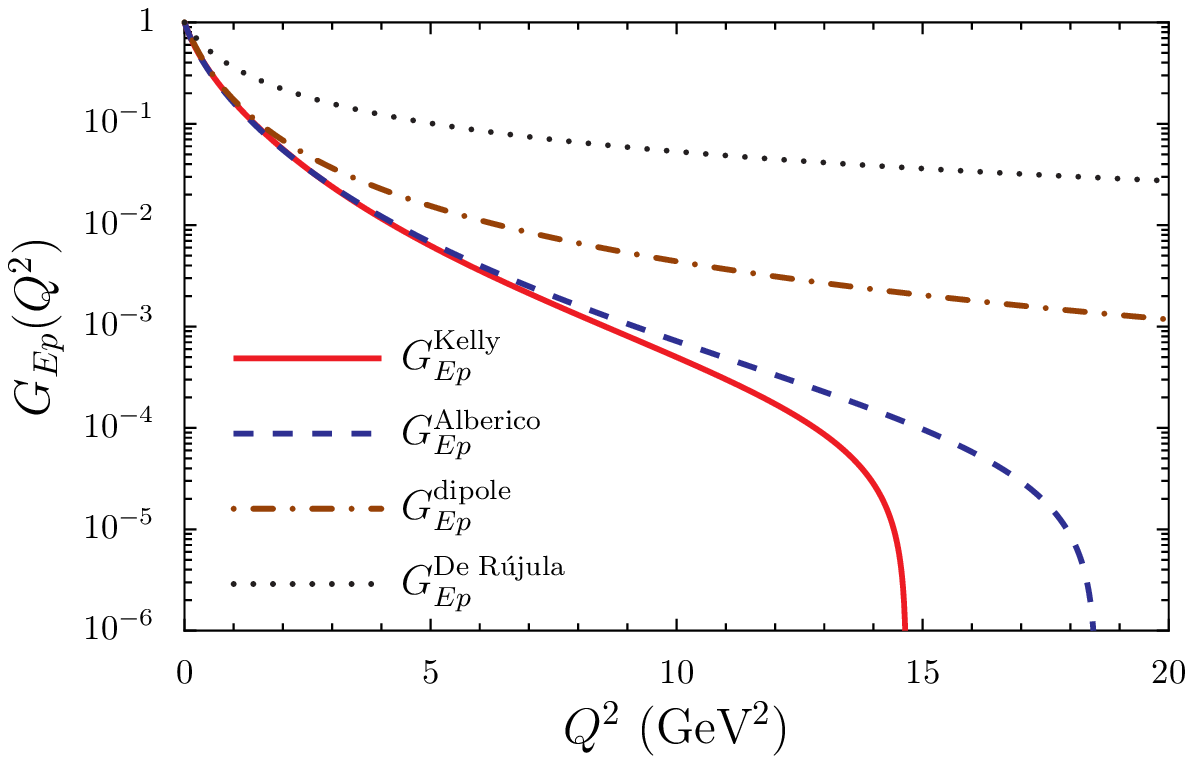} \\
\centering\includegraphics[width=\columnwidth,clip=true,angle=0]{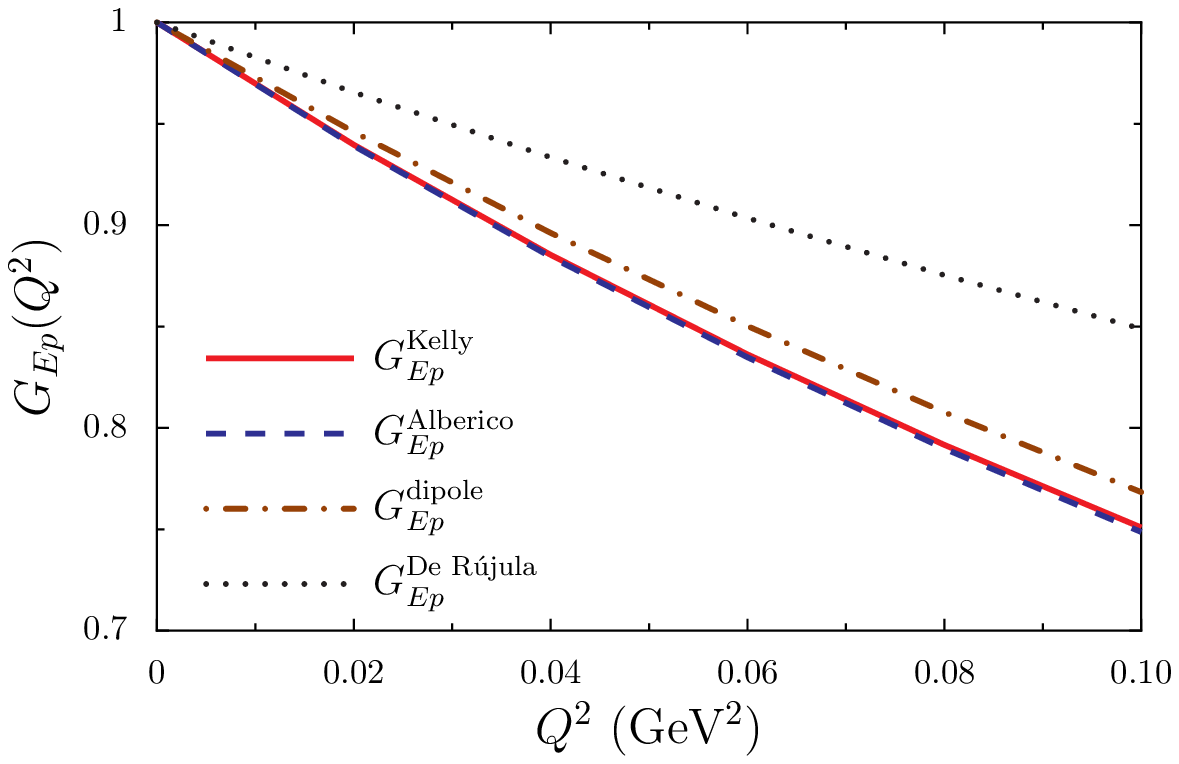}
\caption{The proton $G_{Ep}$ form factor from the parameterizations of Kelly~\cite{kelly}
(solid line), fit II from Alberico \textit{et al.}~\cite{wanda} (dashed line) and
the dipole form factor of Eq.~\eqref{eq:dipole} is given as the dot-dashed line. 
The ``toy model'' of Ref.~\cite{adr}, with $\sin^2\theta = 0.3$, $m=15\,$MeV and $M=750\,$MeV,
is illustrated as the dotted line.}
\label{fig:form}
\end{figure}

Even more recently De~R\'ujula~\cite{adr} pointed out that the discrepancy between
the two results can be removed if the proton's third Zemach moment is very large. He
finds that a value of
\begin{align}
\langle r_p^3 \rangle_{(2)} = 36.6 \pm 6.9\, \text{fm}^3,
\label{zmm}
\end{align}
combines the impressive experimental results from the CODATA compilation and the
recent muonic Lamb shift data in a consistent manner. This result is more than
13 times larger than the experimental extraction of Friar and Sick~\cite{Friar:2005jz}, who
use electron-proton scattering data to determine
\begin{align}
\langle r_p^3 \rangle_{(2)} = 2.71 \pm 0.13\,\text{fm}^3.
\label{eq:friar}
\end{align}
It is noteworthy that the length scale associated with Eq.~\eqref{zmm} is about four times
the proton root-mean-square radius. Such a large value is supported in Ref.~\cite{adr} by 
a ``toy model'' of the proton electric form factor, namely
\begin{align}
G_E^{\text{De R\'ujula}}(Q^2) 
= \frac{1}{D}\left[\frac{c^2\,M^4}{Q^2 + M^2} + \frac{s^2\,m^6}{\left(Q^2+m^2\right)^2}\right],
\label{eq:derujula}
\end{align}
where $s^2 \equiv \sin^2\theta$, $c^2 \equiv \cos^2\theta$ and $D \equiv c^2\,M^2 + s^2\,m^2$.
The CODATA charge radius result and the large Zemach moment of Eq.~\eqref{zmm} can be
accommodated for a range of mass parameters and mixing angle. For $s^2 = 0.3$ the
mass scales are typically $m \simeq 15\,$MeV and $M \simeq 750\,$MeV~\cite{adr}. For
these parameters we have plotted Eq.~\eqref{eq:derujula} as the dotted line
in Fig.~\ref{fig:form}, where it is compared with a dipole form factor and the 
empirical parameterizations from Refs.~\cite{kelly,wanda}.
The severe disagreement between the form factor of Eq.~\eqref{eq:derujula} and the
realistic form factors of Refs.~\cite{kelly,wanda} demonstrate that this
``toy model'' is not a viable representation of the data.

\begin{table*}[btp]
\addtolength{\extrarowheight}{5.0pt}
\begin{ruledtabular}
\begin{tabular}{l|ccccc}
parameterization
& $\langle r_p^3\rangle_{(2)}$ [fm$^3$]
& $\langle r_p^2\rangle$  [fm$^2$]
& $\langle r_p^2\rangle^{1/2}$  [fm]  
& $\langle r_p^4\rangle$  [fm$^4$]
& $\langle r_p^3\rangle_{(2)}^{\text{max}}$ [fm$^3$]\\[0.6ex]
\hline
Dipole (Eq.~\eqref{eq:dipole})          & 2.023 & 0.658 & 0.811 & 1.083 & 3.029\\
Kelly~\cite{kelly}                      & 2.524 & 0.744 & 0.863 & 1.619 & 4.067\\
Alberico \textit{et al.}~\cite{wanda}   & 2.545 & 0.750 & 0.866 & 1.623 & 4.097
\end{tabular}
\end{ruledtabular}
\caption{Values of the third Zemach moment $\langle r_p^3\rangle_{(2)}$,
the mean-square charge radius $\langle r_p^2\rangle$, the charge radius 
$\langle r_p^2\rangle^{1/2}$ and the fourth moment of the charge distribution
$\langle r_p^4\rangle$,
for each of the three proton form factor parameterizations considered. The last
column containing $\langle r_p^3\rangle_{(2)}^{\text{max}}$ represents the upper
bound on the third Zemach moment of the proton given by Eq.~\eqref{eq:zemachbound},
with $\mu = 1\,$GeV.}
\label{tab:res}
\end{table*}

Nevertheless, it is reasonable to ask if the use of a realistic
extraction of $G_{Ep}$ leads to a value of the Zemach radius very
different from that obtained from the typically used dipole form,
namely
\begin{align}
G_E^{\text{dipole}}(Q^2) = \left(1 + \frac{Q^2}{\Lambda^2}\right)^{-2},
\label{eq:dipole}
\end{align}
where $\Lambda^2 = 0.71\,$GeV$^2$.
This form has historical validity in describing early data
and was explicitly assumed in the recent experimental analysis of Pohl \textit{et al.}~\cite{pohl}.
However, several experiments (see for example the review of Ref.~\cite{Perdrisat:2006hj})
have found that $G_{Ep}$ actually falls faster than the dipole. Therefore,
we use two different more recent parametrizations from Refs.~\cite{kelly,wanda}.
These parametrizations take the general form
\begin{align}
\hspace{-1mm}G_E(Q^2) = \frac{1+ \sum_{n=1}^N a_n \tau^n}{1+ \sum_{n=1}^N b_n \tau^n}, \quad \text{where} \quad
\tau \equiv \frac{Q^2}{4\,m_p^2},
\end{align}
with proton mass labeled by $m_p$ and the various coefficients are given in
Refs.~\cite{kelly,wanda}. Note, we use fit II for the Alberico~\textit{et al.}~\cite{wanda} parameterization. 
These empirical results for the proton $G_{Ep}$ form factor are illustrated in
Fig.~\ref{fig:form}.

The third Zemach moment is defined by
\begin{align}
\langle r_p^3\rangle_{(2)}\equiv \int d^3r \;r^3 \rho_2(r),
\label{eq:zemach}
\end{align}
where
\begin{align}
\rho_2(r) &= \int d^3r' \rho_p(\bfr')\rho_p(|\bfr'-\bfr|), \nonumber \\
          &= \int \frac{d^3q}{(2\pi)^3}\ e^{-i\,\bfq\cdot\bfr}\,G_E^2(\bfq^2).
\end{align}
The conventional proton charge density, $\rho_p(r)$, is defined by
\begin{align}
\rho_p(r)\equiv \int \frac{d^3q}{(2\pi)^3}  e^{-i\,\bfq\cdot\bfr}\,G_{Ep}(\bfq^2),
\end{align}
with the charge radius given by
\begin{align}
\langle r_p^2\rangle=\int d^3r \;r^2\rho_p(r) .
\end{align}

The results of our numerical evaluations are presented in Table~\ref{tab:res}.
Observe that the values of $\langle r_p^3\rangle_{(2)}$ are smaller than that of Eq.~\eqref{zmm}
by approximately a factor of 15. However, the agreement with the empirical result of
Ref.~\cite{Friar:2005jz} (see Eq.~\eqref{eq:friar}) is fairly good.
The $G_{Ep}$ parameterizations of Refs.~\cite{kelly,wanda} fall faster with increasing $Q^2$ than the
dipole form factor (see Fig.~\ref{fig:form}), leading to coordinate space distributions of greater extent.
Therefore, the values of the third Zemach moment $\langle r_p^3\rangle_{(2)}$ and
the mean-square radius $\langle r_p^2\rangle$ both increase relative to the
values obtained from the dipole form factor. 

Examination of the integrand in
Eq.~\eqref{eq:zemach} for each of the three form factor models, reveals that 80\%
of the strength for the Zemach moment comes from the domain where $Q^2$ is less than $1\,$GeV$^2$ 
and 50\% from $Q^2$ values less than $0.25\,$GeV$^2$. Therefore, the quoted values
of the Zemach moment given in Table~\ref{tab:res} are insensitive to the model dependent
extrapolation to infinite $Q^2$. In the experimental extraction of the charge radius by
Pohl \textit{et al.}~\cite{pohl}, the third Zemach moment is suppressed relative to
the charge radius by a factor of $\sim 570$, therefore the variation of the 
Zemach moment observed in Table~\ref{tab:res} has negligible impact on the extracted value of 
the charge radius.

\begin{figure}[tbp]
\centering\includegraphics[width=\columnwidth,clip=true,angle=0]{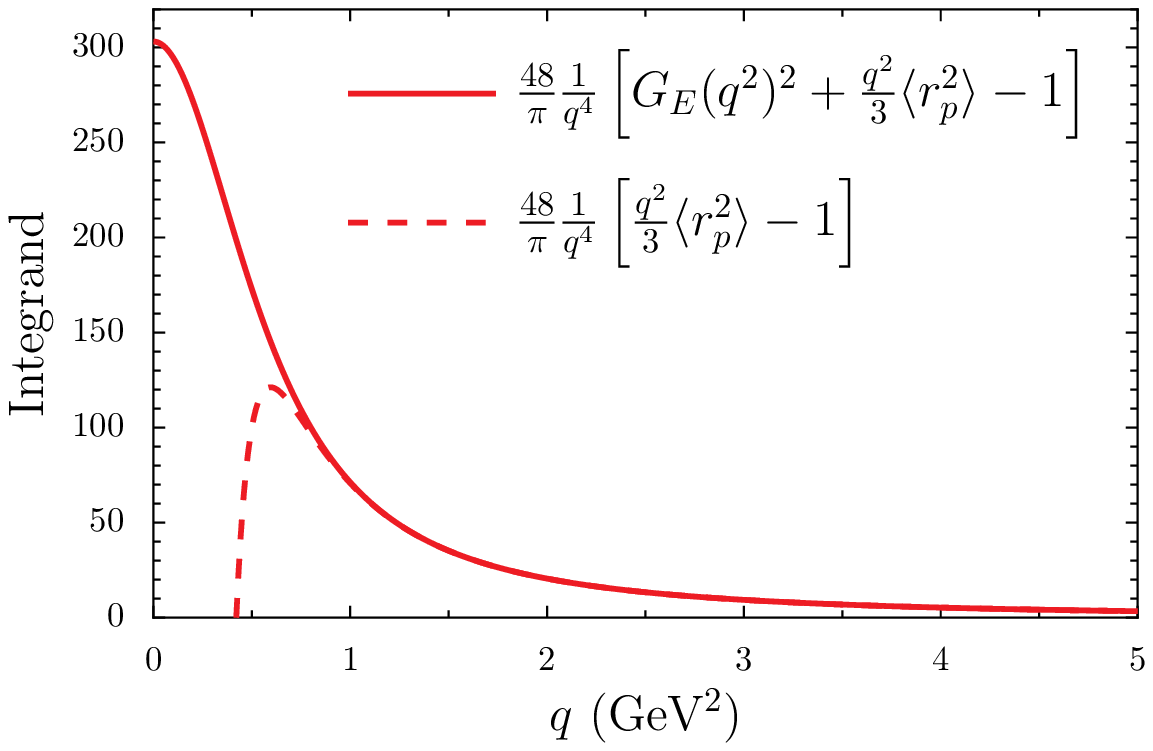}
\caption{Plot of the integrand in Eq.~\eqref{eq:zemachalt} (solid line) and the dominate piece
at moderate to large $q$ given by Eq.~\eqref{eq:zemachlargeq} (dashed line).}
\label{fig:zemach}
\end{figure}

It is possible to obtain a rigorous upper bound on the third Zemach moment
of the proton using the result~\cite{Friar:2005jz}
\begin{align}
\langle r_p^3\rangle_{(2)} = \frac{48}{\pi} \int_0^\infty \frac{dq}{q^4} \
\left[G_{Ep}(q^2)^2 + \frac{q^2}{3} \langle r_p^2 \rangle - 1\right].
\label{eq:zemachalt}
\end{align}
The integrand in Eq.~\eqref{eq:zemachalt} is a monotonically decreasing positive definite 
function on the integration domain, with a maximum at $q=0$ given by
\begin{align}
I(q=0) = \frac{48}{\pi}\left[\frac{1}{60}\langle r_p^4\rangle + \frac{1}{36}\langle r_p^2\rangle^2\right].
\label{eq:zemachupper}
\end{align}
For moderate to large $q$ the integrand in Eq.~\eqref{eq:zemachalt} is completely
saturated by the last two terms, that is
\begin{align}
I \to \frac{48}{\pi}\
\frac{1}{q^4}\left[\frac{q^2}{3} \langle r_p^2 \rangle - 1\right].
\label{eq:zemachlargeq}
\end{align}
Empirically we know that at $q = 1\,$GeV the above expression represents $\sim\!\!99.5$\%
of the integrand in Eq.~\eqref{eq:zemachalt}.
This is illustrated in Fig.~\ref{fig:zemach} for the dipole form factor 
of Eq.~\eqref{eq:dipole}, where we plot the integrand of Eq.~\eqref{eq:zemachalt} and 
Eq.~\eqref{eq:zemachlargeq}. 
The integrand of
Eq.~\eqref{eq:zemachalt} coincides with Eq.~\eqref{eq:zemachlargeq} for $q \gtrsim 1\,$GeV,
this is also the case for the empirical form factors of Refs.~\cite{kelly,wanda} and
the model of De Rujula. 

Therefore, the third Zemach moment has a rigorous upper bound given by
\begin{align}
\langle r_p^3\rangle_{(2)} &\leqslant
\frac{4}{5\,\pi}\left[\langle r_p^4\rangle + \frac{5}{3}\,\langle r_p^2\rangle^2\right]\mu
+ \frac{16}{\pi\,\mu}\left[\langle r_p^2\rangle + \frac{1}{\mu^2}\right].
\label{eq:zemachbound}
\end{align}
where the first term is the area of the rectangle defined by the integrand 
upper bound of Eq.~\eqref{eq:zemachupper} and $0 \leqslant q \leqslant \mu\,$GeV, whereas the second term
results from integrating Eq.~\eqref{eq:zemachlargeq} from $q=\mu\,$GeV to infinity.
For the reasons discussed above we expect that $\mu = 1\,$GeV is a conservative choice for the
evaluation of the upper bound given in Eq.~\eqref{eq:zemachbound}. 
It is therefore apparent
that if a model reproduces the proton charge radius, a large third Zemach moment 
can only occur if $\langle r_p^4\rangle$ is also large. Typical empirical values are
given in the second last column of Table~\ref{tab:res}, whereas the model of De Rujula,
expressed in Eq.~\eqref{eq:derujula}, gives $\langle r_p^4\rangle = 1849\,$fm$^4$.
The upper bound for the third Zemach moment of the proton obtained from 
Eq.~\eqref{eq:zemachbound}, with $\mu = 1\,$GeV, is given in the last column of Table~\ref{tab:res}
where it is labeled by $\langle r_p^3\rangle_{(2)}^{\text{max}}$.

The net result is that the published parametrizations, which take into account a wide
variety of electron scattering data, cannot account for
the value of the third Zemach moment found in Ref.~\cite{adr}. It is
likely that further investigations of low $Q^2$
data will lead to improved  parametrizations for $G_{Ep}$~\cite{arrington}. However,
enhancing the Zemach moment
significantly above the dipole result is extremely unlikely.
We believe that the resolution of the current puzzle
regarding the proton radius, determined from the muon and electron Lamb shifts
in hydrogen, lies in a direction other than that suggested by De R\'{u}jula.

\vspace{2em}
\section*{Acknowledgments}
This research was supported by the United States Department
of Energy. We thank R. Gilman for informing us about Ref.~\cite{adr}
and York Schroeder for bringing to our attention errors in the 
original version of Ref.~\cite{adr}.

\end{document}